# SYNTHESIS AND CHARACTERIZATION OF NANOSTRUCTURED COBALT SULPHIDE DOPED WITH DYSPROSIUM FOR PHOTOVOLTAIC APPLICATION

ENAJITE Oscar[a], M. O. OSIELE[b], OMOYIBO Samuel Emovokeraye[c] and IMONI-OGBE Kingsley[d]

[a]Physics Department, Dennis Osadebay University, Asaba, Delta State, Nigeria.
Email: Oscar.enajite@dou.edu.ng

[b]Physics Department, Delta State University, Abraka, Nigeria
Email: osielemo@delsu.edu.ng

[c]Physics Department, Dennis Osadebay University, Asaba, Delta State, Nigeria.
Email: Samuel.omoyibo@dou.edu.ng

[d]Physics Department, University of Delta, Agbor, Nigeria
Email: kingsley.imoni-ogbe@unidel.edu.ng

## ABSTRACT

The technological world is in search of environmentally friendly methods of generating energy for the use of the ever growing needs of power for sustainable development of visually all facet of life. Solar energy form an environmentally friendly solution for the reduction of global warming. For usage in photovoltaic applications, this work explores the production and characterisation of nanostructured cobalt sulphide (CoS) doped with dysprosium (Dy), a rare earth element. In order to increase CoS's efficiency in solar energy conversion, rare earth doping is used to improve its optical, electrical, and structural characteristics. To maximize performance, precise Dy doping doses were used in the electrochemical synthesis of the nanomaterials. UV-visible spectroscopy were among the characterization methods used to examine the optical and structural characteristics of the doped CoS nanomaterials. Significant alterations in the bandgap and optical absorption behavior are shown by the results, indicating that Dy-Doped CoS nanostructures hold great promise for enhanced photovoltaic performance. This study opens the door for the use of Dy-doped CoS in renewable energy technologies by demonstrating their viability as prospective materials for next-generation solar cells.

**Key Words:** *Cobalt Sulphide, dysprosium, rear-earth element, photovoltaic, Doping, nanostructure*

## INTRODUCTION

The significance of renewable energy sources has increased due to the growing worldwide demand for energy and the negative effects of fossil fuels on the environment. Solar energy is unique among these since it is sustainable and abundant. A key component of the shift to sustainable energy sources is photovoltaic (PV) technology, which directly transforms sunlight into electrical power. For PV systems to be widely adopted, they must be efficient and cost-effective, and improvements in material science are crucial to raising these metrics [1]. Because of their special qualities at the nanoscale, nanostructured materials have become attractive options for improving solar performance. These materials are perfect for enhancing light absorption and charge carrier efficiency because of their larger surface area, quantum confinement effects, and adjustable optical characteristics.
In PV device mobility [2] in particular, it has been demonstrated that doping nanostructured materials with rare earth elements like dysprosium improves their photovoltaic characteristics by adding new electronic states that promote better charge separation and transport [3]. In order to create more robust and effective photovoltaic materials, it is crucial to synthesize and characterize these doped nanostructures, such as cobalt sulphide.

Rare earth elements (REEs) are a group of 17 chemically similar elements, including the lanthanides, scandium, and yttrium, that are widely recognized for their unique electronic, magnetic, and optical properties. These elements have become increasingly important in various high-tech applications, including photovoltaics (PV), due to their ability to enhance the performance and efficiency of materials used in solar cells. Dysprosium (Dy) ,is a prominent REE, PV device mobility [4-7]. In particular, it has been demonstrated that doping nanostructured materials with rare earth elements like as dysprosium improves their photovoltaic characteristics by adding new electronic states that promote better charge separation and transport [8]. In order to create more robust and effective photovoltaic materials, it is crucial to synthesize and characterize these doped nanostructures, such as cobalt sulphide [9]. Since these REEs can greatly improve the optoelectronic characteristics of the host material, the deliberate addition of dysprosium to nanostructured materials like cobalt sulphide (CoS) holds promise for the creation of high-performance solar systems. In order to investigate their potential to increase the efficiency of photovoltaic applications, this study attempts to synthesis and characterize Dysprosium-doped cobalt sulphide nanostructured materials.

Because of its special qualities and ability to increase solar cell efficiency, cobalt sulphide (CoS) has become a material of great promise in the field of photovoltaic technology. Because of its straight bandgap, CoS is a transition metal chalcogenide that can absorb light in the visible portion of the spectrum. For solar applications, its high absorption coefficient, adjustable bandgap, and strong electrical conductivity are very beneficial [10-12]. CoS is also well-known for its simplicity of synthesis and chemical stability, both of which are essential for the creation of long-lasting and reasonably priced solar cells.

Optimizing these characteristics to improve solar cell performance has been the main emphasis of earlier research on CoS-based photovoltaic materials. In dye-sensitized solar cells (DSSCs), studies have shown that CoS can be used as an effective counter electrode material. In these cells, it catalyzes the reduction of triiodide to iodide, increasing the total cell efficiency [13]. In order to enhance charge transfer and light-harvesting capabilities, several studies have investigated the inclusion of CoS into other nanostructures, including nanowires and nanosheets [14]. The investigation of CoS in combination with rare earth elements like dysprosium, however, is still largely unexplored in spite of these developments. Rare earth element doping of CoS may result in the introduction of novel electronic states and improve photovoltaic performance by increasing charge carrier mobility and altering the band structure [15]. In order to investigate the potential of dysprosium-doped cobalt sulfide nanostructured materials for photovoltaic applications, this study will synthesis and characterize these materials.

One of the mainstays of the global movement towards renewable energy is photovoltaic (PV) technology, which directly transforms sunlight into electrical power. The photovoltaic effect, which occurs when specific materials produce electricity when exposed to light, is the fundamental component of PV technology. Due to its abundance and advantageous electrical characteristics, silicon is the most often used material in PV cells; nevertheless, a great deal of research has been done to investigate substitute materials that can provide increased efficiency, lower costs, and the possibility of new uses. Because of their distinct optical and electrical characteristics, which can be adjusted at the nanoscale to maximize light absorption and charge carrier mobility, nanostructured materials have attracted a lot of attention among these substitutes [16].

It is anticipated that the incorporation of these cutting-edge materials into PV systems would enhance device performance and aid in the creation of affordable and environmentally friendly energy solutions. The investigation of rare earth-doped nanostructured materials continues to be a crucial field of study in the effort to better capture solar energy as the need for more effective and adaptable PV technology grows [17].

The fundamental idea behind photovoltaic (PV) cells is the photovoltaic effect, which transforms light energy into electrical energy. When light photons hit a semiconductor material's surface inside a photovoltaic cell, they excite the material's electrons from their valence band to the conduction band, generating pairs of electrons and holes. After moving in the direction of the PV cell's front surface, these free electrons are gathered there and produce an electric current that can be used to generate electricity [18]. The semiconductor material's characteristics, such as its band gap, absorption coefficient, and charge carrier mobility, have a significant impact on how effective this method is [20].

The basic principles of photovoltaic (PV) cells revolve around the conversion of light energy into electrical energy through the photovoltaic effect. When light photons strike the surface of a semiconductor material within the PV cell, they impart their energy to electrons in the material, exciting them from their valence band to the conduction band, thus creating electron-hole pairs. The fundamental idea behind photovoltaic (PV) cells is the photovoltaic effect, which transforms light energy into electrical energy. Electrons in a semiconductor material within the PV cell are excited from their valence band to the conduction band by light photons striking their surface. This process produces electron-hole pairs. After moving in the direction of the PV cell's front surface, these free electrons are gathered there and produce an electric current that can be used to generate electricity [18, 19]. The semiconductor material's characteristics, such as its band gap, absorption coefficient, and charge carrier mobility, have a significant impact on how effective this method is [20]. Research over the years has concentrated on improving the efficiency of photovoltaic cells by creating novel materials with enhanced optical and electrical characteristics. One such substance that has drawn interest is cobalt sulphide (CoS), which is a viable option for photovoltaic applications because of its appropriate band gap, high absorption coefficient, and strong electrical conductivity [21]. Doping CoS with rare earth elements, such dysprosium (Dy), has also been investigated as a way to alter its electrical structure and enhance its photovoltaic capabilities.

Due to their distinct magnetic and optical characteristics, dysprosium and gadolinium may improve light absorption and charge carrier dynamics in CoS, boosting PV cell efficiency [22-23]. According to recent research, doping can also reduce recombination losses and increase the overall charge separation efficiency, which are critical factors for the performance of PV cells [24]. The capacity of a photovoltaic cell to transform light into electrical energy determines its efficiency; the theoretical maximum efficiency is given by the Shockley-Queisser limit;

$$\eta = \frac{P_{out}}{P_{in}} = \frac{V_{oc} \times J_{sc} \times FF}{P_{in}} \qquad (1)$$

where $\eta$ is the efficiency, $P_{out}$ is the output power, $P_{in}$ is the incident light power, $V_{oc}$ is the open-circuit voltage, $J_{sc}$ is the short-circuit current density, and FF is the fill factor.

# ELECTRONIC, OPTICAL AND STRUCTURAL PROPERTIES OF COBALT SULPHIDE

Using the Scherrer equation which relates the crystallite size to the broadening of X-ray diffraction peaks, we have

$$D = \frac{K\lambda}{\beta \cos\theta} \quad (2)$$

where $\lambda$ is the X-ray wavelength, $\beta$ is the full width at half maximum (FWHM) of the peak in radians, $\theta$ is the Bragg angle, $D$ is the crystallite size, and $K$ is the form factor (usually 0.9). It is noted that the size of the crystallite has an inverse relationship with the broadening of the diffraction peaks. We determine the size by rearranging the Scherrer equation as and solve for $D$ we obtained

$$\beta = \frac{K\lambda}{D \cos\theta} \quad (3)$$

A popular approach for figuring out the optical bandgap of semiconductor materials is the Tauc plot method, which is especially useful when examining UV-Vis absorption spectra. This approach is predicated on the idea that the absorption coefficient ($\alpha$) close to the absorption edge fluctuates in accordance with a power law with the photon energy ($h\nu$), which is contingent upon the type of electronic transition [25]. The relationship between the absorption coefficient and photon energy for direct allowed transitions, which are prevalent in semiconductor materials such as cobalt sulphide (CoS), can be written as

$$(\alpha h\nu)^2 = A(h\nu - E_g) \quad (4)$$

where $A$ is a material-dependent constant, $E_g$ is the optical bandgap, $h\nu$ is the photon energy, and $\alpha$ is the absorption coefficient [25]. A graph of $(\alpha h\nu)^2$ against $h\nu$ gives the Tauc plot in practice, and projecting the linear part of the figure to the photon energy axis (x-axis) produces the bandgap $Eg$ which is the location where the extrapolated line crosses the x-axis. This approach is especially pertinent for materials like Dysprosium doped CoS nanostructures since doping may be used to customize the bandgap, which affects the optical characteristics that are essential for photovoltaic applications (Yu & Cardona, 2010). The bandgap, a critical parameter in evaluating the performance of the system, can be easily and accurately estimated using the Tauc plot approach. The development of photovoltaic technology will be greatly impacted by the investigation of the synthesis and characterisation of rare earth element-doped cobalt sulfide nanostructured materials, particularly with Dysprosium (Dy). Rare earth elements are expected to improve the optical, electrical, and thermal characteristics of cobalt sulfide structures, which could result in increased solar cell stability and efficiency. Because of their distinct electronic structures, rare earth metal like Dy can add new energy levels to the cobalt sulfide bandgap, improving charge carrier dynamics and increasing light absorption. The development of more ecological, economical, and efficient next-generation solar materials depends heavily on study like this. Furthermore, being aware of how doping affects the structural and functionality of nanomaterial for photovoltaics applications. Cobalt sulfide's characteristics may open the door for more widespread uses of these materials in other industries, such as catalysis and sensors. As a result, our research advances both basic scientific understanding and the useful creation of cutting-edge materials for applications involving renewable energy.

## MATERIALS AND METHOD

Analytically grade Cobalt (II) nitrate hexahydrate (Co $(NO_3)_2.6H_2O$), Thiourea ($CH_4N_2S$), dysprosium oxide ($DyO_3$), and hydrochloric acid (HCl) were the chemicals used for the deposition of dysprosium doped cobalt sulphide (Dy-doped CoS) nanostructured material. These chemicals were obtained and used straight away without additional purification. Beakers, distilled water, magnetic stirrer, voltage regulator, multimeter, ammeter, carbon electrode, fluorine electrode, working electrode, heater mantle, electronic weighing balance, masking tape, stopwatch, UV/visible Spectrophotometer (UV-1800), etc. are additional materials used.

## SUBSTRATE CLEANING

The deposition of thin films relies heavily on substrate cleaning. Commercially obtained glass slides were used for the deposition. After cleaning the glass substrates with a mild soap solution, they were degreased using acetone, etched with 5% HCl for 30 minutes, ultrasonically cleaned with distilled water, and air-dried before deposition.

## METHOD OF PREPARATION:

Before putting the material compounds into the beakers, we measured their respective masses using the electronic scale. 5 ml of HCl was used to dissolve the dysprosium oxide ($DyO_3$), after some minutes; 50 ml of water was added and stirred. Cobalt (II) nitrate hexahydrate ($Co(NO_3)_2.6H_2O$) and Thiourea ($CH_4N_2S$) compounds of 0.1 M was dissolved in 100 ml of water to form pink and white solutions and were then stirred using the magnetic stirrer for five minutes.

## METHOD OF DEPOSITION

Prior to the nanostructured material being deposited, the glass substrates coated with FTO were weighed. After weighing, the conducting sides of the FTO glass substrate were identified using a multimeter. During deposition, the target materials were put into the beakers. The substrate was removed from the glass container using forceps, positioned between the working electrodes with its conducting side facing the fluorine electrode, and then inserted into the target material. A digital multimeter was used to measure the potential drop across the thin film for 10 seconds, and a sensitive ammeter was used to measure the current flowing through the sample. A steady 10 V was maintained in the power supply. The thin coatings of cobalt sulfide that were still forming were initially applied at ambient temperature to the first glass substrate coated with FTO, the dysprosium oxide (DyO3) dopant was adjusted between 0.01 and 0.03 M for the final three substrates. Through the electrolysis process, the electrochemical approach was used to accomplish the deposition. To eliminate internal tensions, the samples were once more placed in the annealing process and heated for 30 minutes following the deposition. A UV-1800 spectrophotometer was used to analyze the optical absorption of the film produced on FTO coated glass (substrate) in the wavelength range of 300–1100 nm, and an X-ray diffractometer was used to study the structure.

Table 1: Variations of Concentration

| Material (Molarity) | (Co(NO$_3$)$_2$.6H$_2$O) (ml) | (C$_2$H$_5$NS) (ml) | DyO$_3$ (ml) | Time (Sec) | Voltage (V) |
|---|---|---|---|---|---|
| CoS | 20 | 20 | 0 | 10 | 10 |
| CoS/Dy 0.01 mol | 15 | 15 | 5 | 10 | 10 |
| CoS/Dy 0.02 mol | 15 | 15 | 5 | 10 | 10 |
| CoS/Dy 0.03 mol | 15 | 15 | 5 | 10 | 10 |

## RESULTS AND DISCUSSIONS

### Structural analysis of CoS and Dy/CoS material

Figure 1 depicts the X-ray diffraction spectrum of CoS and CoS/Dy material at various dysprosium dopant concentrations of 0.01 mol, 0.02 mol, and 0.03 mol. The films are comprised of polycrystalline materials. At orientation (111), the film showed a noticeable peak that corresponded to a 2theta value of 26.611°. At diffraction planes (111), (112), (200), and (311), which correspond to 2theta values of 26.611°, 33.895°, 38.112°, and 51.831°, respectively, a flawless diffraction peak with a hexagonal phase was found. A rise in peak intensity at higher 2theta degree values indicates the presence of a crystal lattice. Because of their intrinsic qualities, polycrystalline materials are more effective for solar and photovoltaic construction. Higher film thickness and dysprosium dopant concentration may be the cause of the spectrum's higher peaks, which provide a greater surface area for solar cell and photovoltaic activity. The average crystallite size of the films, spectral characteristics, and computed crystallite or grain sizes are listed in Table 2.

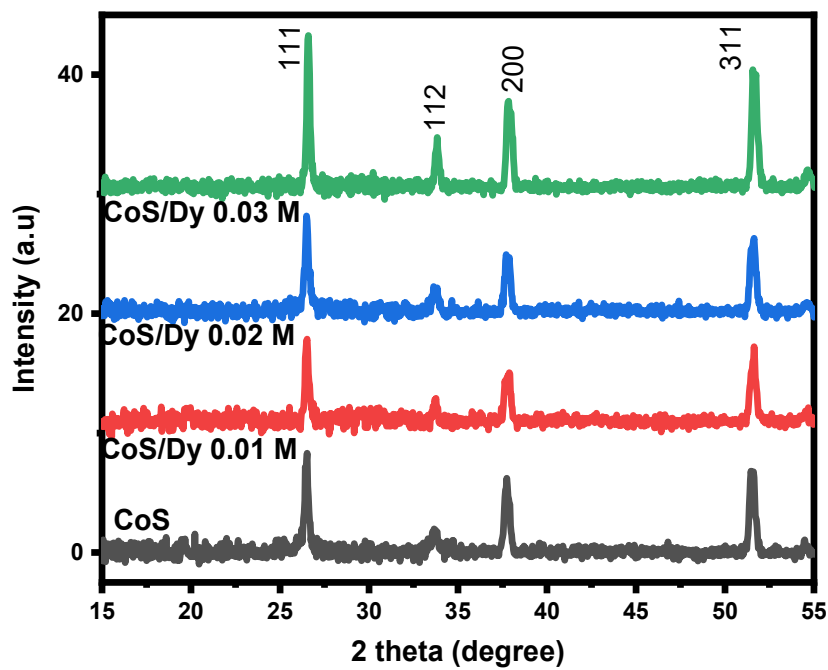

Figure 1: XRD pattern of CoS and CoS doped with Dy material.

**Table 2: CoS and CoS doped with Dy material Structural values.**

| Films | 2θ (Degree) | d (spacing) Å | (Å) | (β) | (hkl) | (D) nm | σ lines/m² x 10¹⁴ |
|---|---|---|---|---|---|---|---|
| CoS pristine | 26.242 | 3.395 | 5.881 | 0.1372 | 111 | 1.038 | 2.824 |
|  | 33.411 | 2.681 | 5.362 | 0.1395 | 112 | 1.038 | 2.824 |
|  | 37.431 | 2.402 | 4.804 | 0.1342 | 200 | 1.091 | 2.555 |
|  | 51.365 | 1.778 | 3.977 | 0.1348 | 311 | 1.142 | 2.334 |
| CoS/Dy 0.01 M | 26.611 | 3.349 | 5.801 | 0.0686 | 111 | 2.078 | 7.050 |
|  | 33.895 | 2.644 | 5.288 | 0.0684 | 112 | 2.120 | 6.772 |
|  | 38.112 | 2.360 | 4.721 | 0.0683 | 200 | 2.149 | 6.593 |
|  | 51.831 | 1.763 | 3.943 | 0.0681 | 311 | 2.265 | 5.935 |
| CoS/Dy 0.02 M | 26.611 | 3.349 | 5.801 | 0.0343 | 111 | 4.157 | 1.762 |
|  | 33.895 | 2.644 | 5.288 | 0.0347 | 112 | 4.180 | 1.743 |
|  | 38.112 | 2.360 | 4.721 | 0.0349 | 200 | 4.206 | 1.721 |
|  | 51.831 | 1.763 | 3.943 | 0.0352 | 311 | 4.382 | 1.585 |
| CoS/Dy 0.03 M | 26.611 | 3.349 | 5.801 | 0.0359 | 111 | 3.971 | 1.930 |
|  | 33.895 | 2.644 | 5.288 | 0.0357 | 112 | 4.063 | 1.844 |
|  | 38.112 | 2.360 | 4.721 | 0.0353 | 200 | 4.158 | 1.761 |
|  | 51.831 | 1.763 | 3.943 | 0.0352 | 311 | 4.382 | 1.585 |

**Electrical study**

Table 3 and Figure 2 display the conductivity and resistivity of CoS and CoS/Dy materials synthesized at different dopant concentrations. The film resistivity dropped from 99.32 to 62.23 nm, while the materials' thickness decreased from 114.98 to 103.73 nm. Conductivity increased from 1.00 to 1.60 S/m as a result. Dysprosium dopant concentrations of 0.01 mol, 0.02 mol, and 0.03 mol were applied to the materials. The film resistivity decreased from 99.32 to 68.00, 65.95 and 62.23 respectively and the materials' thickness also decreased from 114.98 to 108.99 nm, 104.25 and 103.73nm respectively. The conductivity increased from 1.00 to 1.47, 1.51, and 160 S/m respectively as a result. Dysprosium dopant concentrations of 0.01 mol, 0.02 mol, and 0.03 mol were used to deposit the materials. Due of their low resistance and high conductivity, the films produced will be suitable for solar and photovoltaic applications. Low resistance and high conductivity increase and decrease with layer thickness, as seen in Figures 2 and 3. The link is shown in a nonlinear graph in Figures 2 and 3 between resistivity, conductivity, and the level of dysprosium dopants, with resistivity and conductivity fluctuating as dopant levels increase. The graph also shows resistance and conductivity versus thickness.

**Table 3: Electrical properties of the material**

| Material (Films) | Thickness, t (nm) | Resistivity, $\rho$ ($\Omega$.cm) X $10^4$ | Conductivity, $\sigma$ (S/m) X $10^2$ |
|---|---|---|---|
| CoS | 114.98 | 99.32 | 1.00 |
| CoS/Dy 0.01 M | 108.99 | 68.00 | 1.47 |
| CoS/Dy 0.02 M | 104.25 | 65.95 | 1.51 |
| CoS/Dy 0.03 M | 103.73 | 62.23 | 1.60 |

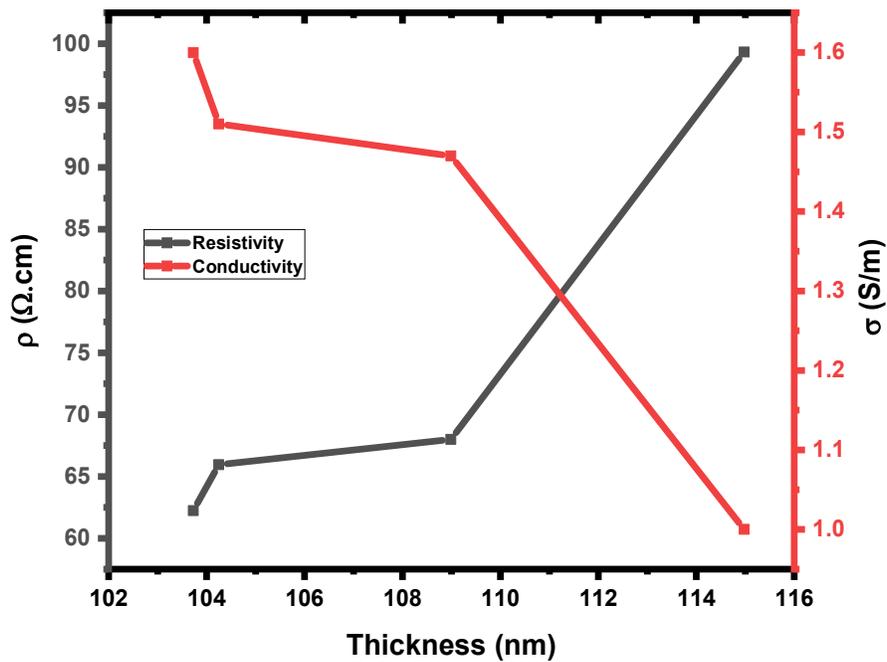

Figure 2: plot of resistivity and conductivity Vs thickness

## Optical analysis of CoS and Dy/CoS material

Figure 3 depicts the absorbance spectra of CoS and CoS/Dy materials produced with varying dysprosium dopant concentrations. As the wavelength of the incident light radiation increases, the CoS's absorbance drops. In the case of dysprosium dopant material, the absorbance of the films increases with increasing wavelength. The CoS film has the highest absorption in both spectral areas. As the concentration of dysprosium grows in the UV area, the film thickness decreases, but marginally increases in the visible range. The doped material has moderate absorbance in both spectra, making it suited for industrial photovoltaic panel production.

Figure 4 depicts the transmittance spectra of CoS and CoS/Dy materials produced with varying dysprosium dopant concentrations. As the wavelength of the incident light radiation increases, the CoS's transmittance drops. In the case of dysprosium dopant material, the films' transmittance increases as the material's wavelength increases. The CoS film has the lowest transmittance in both spectra. As the concentration of dysprosium grows in the UV area, the film thickness decreases, but marginally increases in the visible range. The doped material has a higher transmittance in both spectrum regions, making it suited for the industrial construction of photovoltaic solar panels.

Figure 5 depicts the reflectance spectra of CoS and CoS/Dy materials produced with varying dysprosium dopant concentrations. As the wavelength of the incident light radiation increases, the CoS's reflectance drops. In the case of dysprosium dopant material, the reflectance of the films increases with increasing wavelength. The CoS film has the highest reflectivity in both spectra. As the concentration of dysprosium grows in the UV area, the film20 thickness decreases, but marginally increases in the visible range. The doped material has moderate reflectance in both bands, making it suited for industrial photovoltaic panel production.

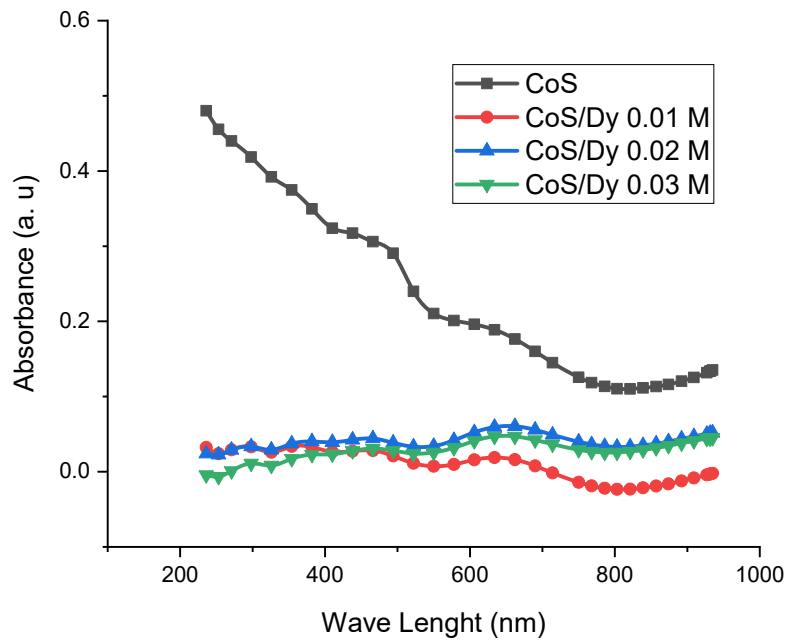

Figure 3: plot of absorbance as a function of wavelength for CoS and CoS dopped with different concentration of Dy

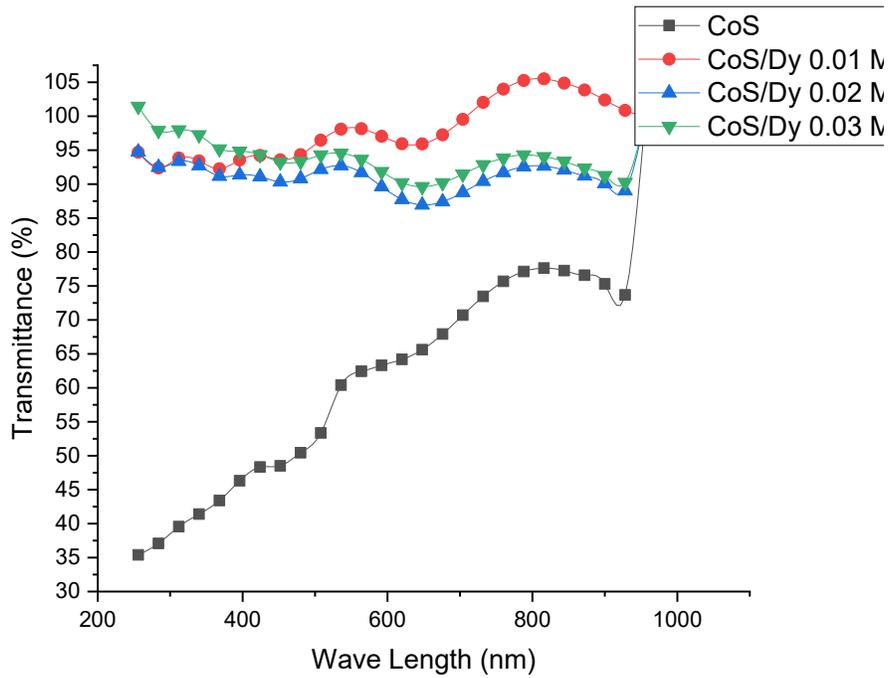

Figure 4 plot of transmittance as a function of wavelength for CoS and CoS dopped with different concentration of Dy

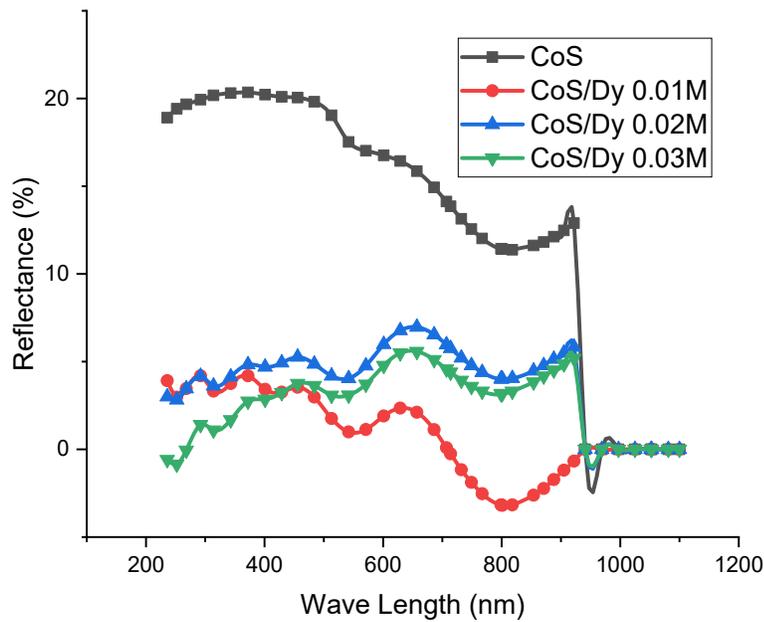

Figure 5: plot of reflectance as a function of wavelength for CoS and CoS dopped with different concentration of Dy

Figure 6(a,b) shows the relationship between (αhυ)² and hυ for CoS and CoS/Dy materials with varying dopant concentrations. The energy band gaps for CoS and CoS/Dy materials at various dopant concentrations were determined from Figure 6 by extrapolating the straight portion of the graph down to the hv axes at (αhυ)² = 0. The energy band gaps of the deposited films were found to be between 1.50 and 2.98 eV. The introduction of dysprosium as a dopant raised the energy band gap, Eg, which increased further with dysprosium concentration. In conclusion, dysprosium dopant was discovered to improve the energy band gap of CoS. The $E_g$ range observed in this work is suitable for the absorber layer in solar cells, as it absorbs all solar energy radiation larger than 1.50 eV and theoretically fits the sun's maximum spectrum (band gap ~1.5 eV).

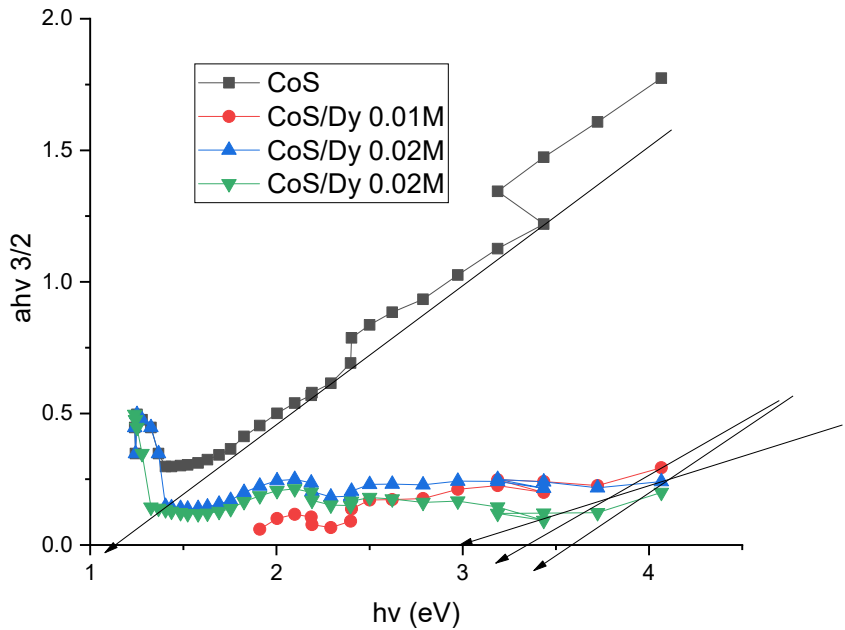

Figure 6a: A plot of *(αhv)²* and *hυ* (eV) for CoS and CoS/Dy materials

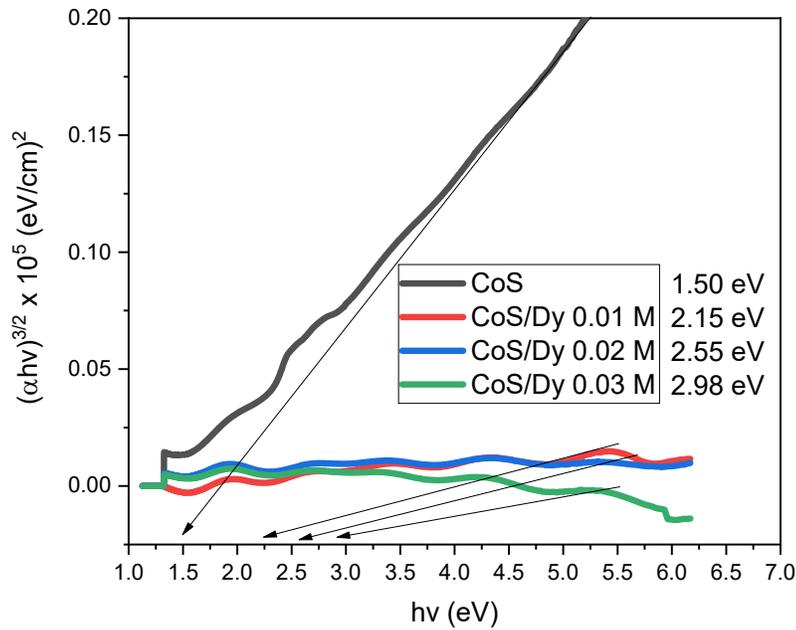

Figure 6b: A plot of *(αhʋ)*x10⁵(eV/cm)² and *hʋ* (eV) for CoS and CoS/Dy materials

Figure 7 shows the refractive indexes of CoS and CoS/Dy materials at different dopant concentrations. The refractive index determines how light bends through things such as water or glass. It compares the speed of light in a vacuum to its speed in a given medium. The refractive index (n) is calculated by dividing the speed of light in a vacuum by that of light in the medium. The refractive index is dimensionless due to the velocity ratio. A higher photon energy yields a higher refractive index. Dysprosium enhances the refractive index of CoS materials. Higher doping concentrations (0.02 mol) have a particularly pronounced effect. Dysprosium ions are more polarizable than Co ions, resulting in an enhanced refractive index. Increased RI can help applications such as optical waveguides, lenses, and sensors. Excessive dysprosium doping can cause the formation of undesired secondary phases, which can have an effect on the materials' characteristics. The presence of dysprosium has major implications.

Figure 8 is graph of optical conductivity as a function of photon energy for CoS and CoS dopped with different concentration of Dysprosium, the plot shows that varied molar concentrations (0.01 to 0.03 mol) of dysprosium have a significant influence on CoS optical conductivity. This effect is achieved by increasing the energy levels of CoS, resulting in increased light absorption and scattering. The concentration of dysprosium dopant affects the optical conductivity, resulting in varying degrees of influence. The conductivity increase is mild at low concentrations (0.01 mol), but much greater at higher values (0.03 mol). Dysprosium dopant atoms produce new defect states in the CoS bandgap, which increases optical conductivity. The existence of defect states increases electron mobility, which leads to higher conductivity. Dysprosium doping alters the dielectric characteristics of CoS material.

Figure 9 shows the extinction coefficient for CoS material. The impact varies depending on the dopant concentration and how it is deposited. Adding a little quantity of dysprosium (0.01 mol) improves the extinction coefficient of CoS, resulting in greater light absorption. When the concentration is increased (0.02 mol), the effect shifts, resulting in a lower extinction coefficient. Dysprosium doping has a complex effect on the extinction coefficient because it brings new energy levels into CoS bandgap, altering its electrical structure and influencing light absorption. The incorporation of dysprosium into CoS materials opens up new possibilities for their use in solar cells, optoelectronic devices, and sensors. These materials' extinction coefficients can be modified, allowing them to be customized for specific light absorption applications.

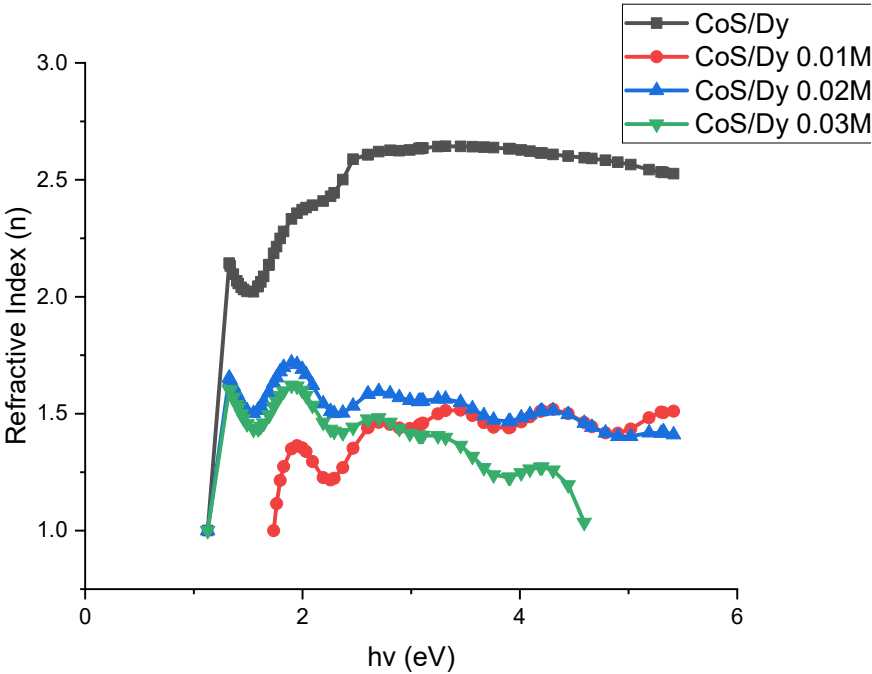

Figure 7: plot of refractive index as a function of photon energy for CoS and CoS dopped with different concentration of Dy

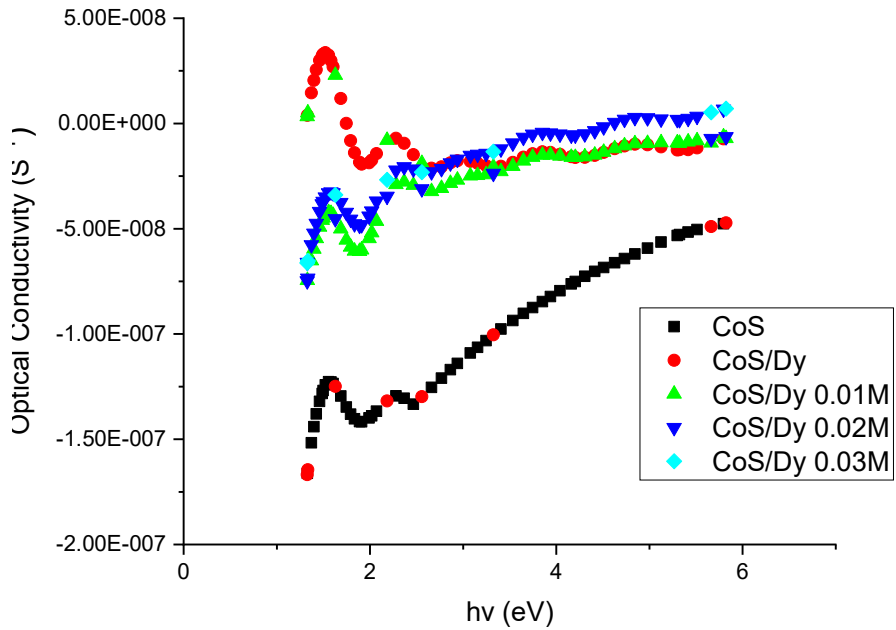

Figure 8: plot of optical conductivity as a function of photon energy for CoS and CoS dopped with different concentration of Dy

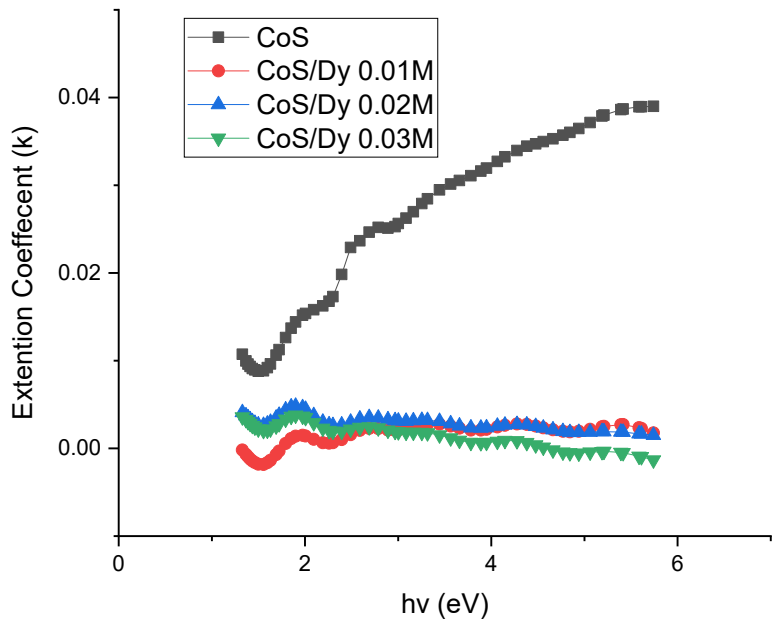

Figure 9: plot of extinction coefficient as a function of photon energy for CoS and CoS doped with different concentration of Dy

The concentration and molarity of dysprosium have a considerable effect on the real and imaginary dielectric constants in Figures 10 and 4.11. Increasing the dysprosium concentration from 0.01 to 0.03 mol significantly increases the actual dielectric constant ($\varepsilon'$). The rise in material polarization is ascribed to defect dipoles and dysprosium ions, which cause enhancement. In contrast, the hypothetical dielectric constant (ε″) declines with increasing dysprosium concentration. The material has lower dielectric loss and higher electrical conductivity.

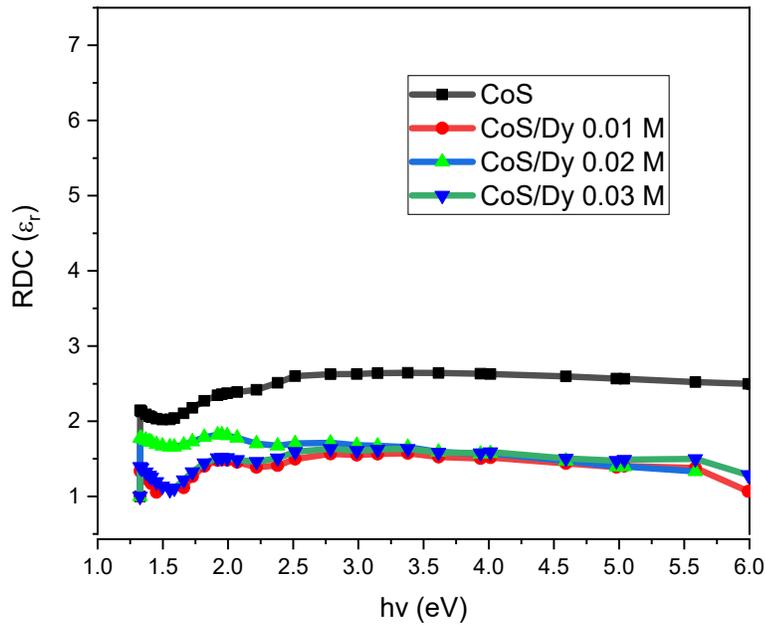

Figure 10: plot of real dielectric constant as a function of photon energy fosr CoS and CoS dopped with different concentration of Dy

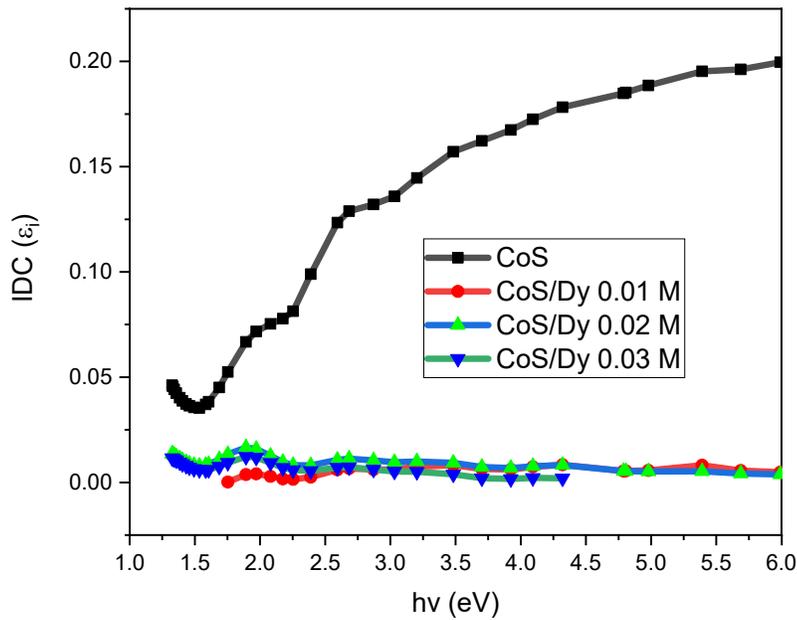

Figure 11: plot of imaginary dielectric constant as a function of photon energy for CoS and CoS dopped with different concentration of Dy

# CONCLUSION

The paper presents a comprehensive summary of the deposition, characterization, and evaluation of dysprosium and gadolinium-doped cobalt sulphide (CoS) nanostructured materials, focusing on the structural analysis of the CoS and dysprosium doped CoS materials which revealed a films of polycrystalline, with distinct peaks corresponding to various diffraction planes, indicating successful doping and the formation of a hexagonal phase. For dysprosium-doped CoS (CoS/Dy), the diffraction patterns exhibited stronger peaks with increased dopant concentration, suggesting improved crystallinity and film quality. Electrical measurements demonstrated that dysprosium doping influenced the resistivity and conductivity of the CoS films. Dysprosium doping led to decreased resistivity and increased conductivity, particularly at higher dopant concentrations, indicating enhanced electrical performance.

Optical analysis revealed variations in absorbance, transmittance, and reflectance with different dopant concentrations. Dysprosium-doped films exhibited increased absorbance and reflectance, making them suitable for photovoltaic applications. The energy band gaps of the doped films were significantly affected by the type and concentration of the dopant, with dysprosium contributing to a broader range of optical energy absorption.

This study demonstrates the potential of dysprosium doping to enhance the properties of cobalt sulphide (CoS) nanostructured materials for photovoltaic applications. By employing electrochemical deposition techniques, we successfully synthesized CoS films doped with dysprosium achieving significant modifications in their structural, electrical, and optical characteristics.

The structural analysis confirmed the formation of high-quality polycrystalline films with improved crystallinity for dysprosium doping, and enhanced film thickness. These modifications were reflected in the electrical properties, where dysprosium-doped CoS exhibited increased conductivity and reduced resistivity.

Optical studies revealed that dysprosium-doped films had superior absorbance and reflectance properties, making them suitable for applications requiring high light absorption. Dysprosium as a dopants significantly altered the energy bandgaps of the CoS films, demonstrating the versatility of doping in tuning material properties for desired applications. The research has led to some significant findings regarding the synthesis and application of dysprosium-doped cobalt sulphide nanostructures effectively tailors the material's performance for photovoltaic applications. These doped nanostructured materials show promise for advancing solar cell technology and other optoelectronic devices, with each dopant providing distinct benefits that can be leveraged on the specific requirements of the application. Future research should explore the integration of these materials into prototype devices to validate their performance in practical scenarios and further optimize their properties for commercial use. The study shows how doping with dysprosium affects the characteristics and functionality of cobalt sulphide nanostructures, emphasizing the advantages of these rare earth elements in improving electrical, optical, and structural characteristics.